\input psfig.sty
\def\pn{{\par\noindent}}

\font\small = cmr7
\def\thn{{\thinspace}}
\def\sc{\scriptstyle}

\def\={\thn\thn=\thn\thn}

\def\tgs{{\thn \rlap{\raise 0.5ex\hbox{$\sc  {>}$}}{\lower 0.3ex\hbox{$\sc  {\sim}$}} \thn }}
\def\tls{{\thn \rlap{\raise 0.5ex\hbox{$\sc  {<}$}}{\lower 0.3ex\hbox{$\sc  {\sim}$}} \thn }}
\def\tll{{\raise 0.3ex\hbox{$\sc  {\thn \ll \thn }$}}}
\def\tgg{{\raise 0.3ex\hbox{$\sc  {\thn \gg \thn }$}}}
\def\tle{{\raise 0.3ex\hbox{$\sc  {\thn \le \thn }$}}}
\def\tge{{\raise 0.3ex\hbox{$\sc  {\thn \ge \thn }$}}}
\def\tl{{\raise 0.3ex\hbox{$\sc  {\thn < \thn }$}}}
\def\tg{{\raise 0.3ex\hbox{$\sc  {\thn > \thn }$}}}
\def\ts{{\raise 0.3ex\hbox{$\sc  {\thn \sim \thn }$}}}
\def\vh{V_{\rm H}}

\def\do{{\tt "}}
\def\tp{{\raise 0.3ex\hbox{\small +}}}

\def\sep{{\par \noindent \hangindent=15pt \hangafter=1}}
\def\deg{{^\circ}}

\def\z{\ \ \ }

\def\asec{{^{\prime\prime}}}
%
%
%

\ifx\mnmacrosloaded\undefined \input mn\fi

%

\newif\ifAMStwofonts

\ifCUPmtplainloaded \else
  \NewTextAlphabet{textbfit} {cmbxti10} {}
  \NewTextAlphabet{textbfss} {cmssbx10} {}
  \NewMathAlphabet{mathbfit} {cmbxti10} {} 
  \NewMathAlphabet{mathbfss} {cmssbx10} {} 
  \ifAMStwofonts
    \NewSymbolFont{upmath} {eurm10}
    \NewSymbolFont{AMSa} {msam10}
    \NewMathSymbol{\upi}     {0}{upmath}{19}
    \NewMathSymbol{\umu}     {0}{upmath}{16}
    \NewMathSymbol{\upartial}{0}{upmath}{40}
    \NewMathSymbol{\leqslant}{3}{AMSa}{36}
    \NewMathSymbol{\geqslant}{3}{AMSa}{3E}

     \let\le=\leqslant
     \let\ge=\geqslant
  \else
    \def\umu{\mu}
    \def\upi{\pi}
    \def\upartial{\partial}
  \fi
\fi


\pageoffset{-2.5pc}{0pc}

\loadboldmathnames



\pagerange{1--7}    
\pubyear{1989}
\volume{226}

\begintopmatter  

\title{A Catalog of Multiplicity among Bright Stellar Systems}
\author{P. P. Eggleton$^1$ and A. A. Tokovinin$^2$}
\affiliation{$^1$Lawrence Livermore National Laboratory, 7000 East Ave, Livermore, CA94551, USA}
\affiliation{$^2$Cerro Tololo Inter-American Observatory, Casilla 603, La Serena, Chile} 

\shortauthor{P. P. Eggleton and A. A. Tokovinin}
\shorttitle{Bright Multiple Systems}


\acceptedline{Accepted .... . Received ....}

\abstract {We consider the multiplicity of stellar systems with (combined) magnitude
brighter than 6.00 in Hipparcos magnitudes. We identify 4559 such bright
systems (including the Sun), and the frequencies of multiplicities 
1, 2, \dots, 7 are found
to be 2718, 1437, 285, 86, 20, 11 and 2.  We discuss the uncertainties,
which are substantial. We also consider the distributions of periods 
of orbits and sub-orbits. We note that for the even more restricted set 
of 478 systems with $\vh\tle 4.00$ the proportions of higher multiples up 
to sextuple are progressively larger (213, 179, 54, 19, 8, 5), suggesting 
substantial incompleteness in even the reasonably well-studied larger sample.
\par This sample can be seen as relatively thoroughly studied for
multiplicity, and reasonably representative of stars more
massive than the Sun. But the restriction to $\vh\tle 6$ means that our 
sample contains hardly any systems where {\it all} components are low-mass 
main-sequence stars (K or M).
\par Data on multiplicity is important as a constraint on (a) the 
star-formation problem, (b) the problem of the evolution of the Galactic 
stellar population, and (c) the interaction of dynamics and evolution
through the effect of Kozai cycles. We discuss these topics briefly.}

\maketitle  

\section{Introduction}
\par The statistics of stellar multiplicity, i.e. the number of components,
the distribution of periods, mass ratios etc., is poorly known, especially
for multiplicities higher than two. Yet it is important in many respects,
for instance as a characteristic of the star-formation process, and as an 
initial
condition for stellar evolution. Our goal here is to determine the observed
multiplicity of a reasonably well defined, well observed, and moderately 
large bright star sample. This is probably the best-studied sample of its
size, of stars sufficiently massive to be important for Galactic evolution.
The disadvantages, as well as advantages, of using a magnitude-limited rather 
than distance-limited sample are discussed below. In a forthcoming paper the 
observed statistics of multiple stars will be compared with a simulated sample, 
including selection effects.

\par The Bright Star Catalogue (BSC: Hoffleit \& Jaschek 1983) is a fundamental
resource when considering the stellar population of the Galaxy, or at any
rate the nearer parts of the Galaxy.  It lists multiplicities, but these
are often visual multiplicities that may include line-of-sight, or `optical',
coincidences. Although roughly limited to magnitude 6.5, it is not entirely
complete to this magnitude, and also includes several fainter stars. The
Multiple Star Catalog (MSC: Tokovinin 1997) carefully identifies many
multiple systems, but restricts itself to multiplicity $\ge 3$. The Hipparcos
Catalog (HIP: Perryman et al. 1997) has useful data such as parallaxes and 
proper motions that can help to distinguish optical from physical systems. 
The MSC  provides  Johnson $V$-magnitudes  of the  brightest
`resolved' companion, rather than  combined magnitudes for the whole
systems. The MSC is  constantly updated, and at the  time of  writing it
contains 348 systems with multiplicity $\tge 3$ and $V\tle 6$.

\par Before counting bright multiple systems, it is necessary to define
both `system' and `bright'. At a first glance, one would like to define
a system as a collection of stars that are gravitationally bound to each other,
but not to neighboring systems. Unfortunately, because gravity is a long-range
force, it is difficult if not impossible to draw a clear boundary. The entire
Galaxy can be viewed as a single system. Intermediate between the Galactic
scale and the scale of individual stars are Galactic clusters, globular
clusters (at least two of which are bright enough in total to qualify), and
various collections of stars, such as groups and associations, which
might qualify as `systems'. We discuss this issue in Section 2.

\par For `bright', we choose the HIP magnitude scale, as being reasonably
homogeneous. But the issue of magnitude is somewhat complicated by the
fact that we wish to use the {\it combined} magnitude of the system. The
348 systems of the MSC referred to above involve Johnson rather than Hipparcos 
magnitudes, and when adjusted for Hipparcos magnitudes the number is 330.

\par Three comparable 7th magnitude components can make a 6th magnitude
system. The obvious alternative, that we might use the magnitude of
the brightest component, seems to us to introduce an unnecessary extra
uncertainty, since for any observing technique there will be systems
that are marginally resolved, and the individual magnitudes will be
less certain than the combined one. However, logic then dictates that
we combine magnitudes even for systems that {\it are} well resolved. Some
bright systems have components several hundred seconds apart. A part
of our purpose is to compare the observed distribution of multiples
with a theoretical model, and in the latter it is obviously most
rational simply to combine the magnitudes of all the components.

\par Our main reason for preferring the Hipparcos magnitude scale is that
it averages the magnitude of variable stars in a logical and systematic way.
The larger-amplitude variables, such as $\delta$ Cep and Mira variables, are
often listed in catalogs under a magnitude that is not a systematic average,
and which may differ from the average by half a magnitude or more. 

\par Our main aim in this paper (Section 3) is to determine as well as 
possible the {\it observed} distribution of multiplicity, along with the 
distribution of periods, for a reasonably large and yet reasonably well-studied
and complete sample. The fact that the sample is magnitude-limited rather than
distance-limited makes it unrepresentative of the lower end of the
mass spectrum, but is we believe compensated by the fact that there is in
effect much greater S/N. Somewhat by coincidence, this sample is in practice
rather well representative of those stars massive enough to have significant
evolution in the course of a Hubble time. A distance-limited sample would
have to go out to $\ts 250\thn$pc to include {\it any} O stars; it would then
include over $10^7$ stars, the vast majority of which would not be at all
well-studied as to multiplicity. In addition, a considerable majority
would be too low-mass to have significant evolution within a Hubble time.

\section{Multiplicity}
\par The great majority of multiple systems are `hierarchical', with
say a wide `binary' containing two closer `binaries'. A few hierarchies 
can be as many as four deep, as in $\nu$ Sco (HR~6026/7) which is one
of the two septuple systems in the bright sample.
Since there is usually a factor of $\ts 10^2 - 10^3$ difference in
separation, from one level of hierarchy to the next, such four-deep
systems are likely to be rare. This large factor accounts for the
stability of hierarchical multiples, although in principle, and if
the orbits and sub-orbits are roughly coplanar, a factor of only
three or four allows long-term stability. However non-hierarchical
systems exist, in small numbers. They can be expected to be young.
and to break up in a few million years as a result of gravitational 
interaction, and presumably that is why they are rare.

\par But there is a substantial grey area where non-hierar\-chi\-cal multiples
can overlap with clusters. Perhaps the entire Pleiades cluster should
be seen as one non-hierarchical system. We prefer to see it as
several independent hierarchical systems, but it is not clear that
there is any sensible criterion that would compel us to do this.

\par Note that we use the word `system', and even the words `multiple
system', to include the possibility of multiplicity one, i.e. single
stars. On the whole we shall avoid the word `star' because this is
often used ambiguously, to mean either a system if the individual
components are not readily distinguishable, or a component of a
system if they are. In this paper we count systems, and within systems
we count components, so far as we are able. We should add that we
do not include planets as components, although we note systems that
contain planets. This of course introduces another gray area, since
it is not clear where stars stop and planets begin.

\par The prototype of a non-hierarchical system is the Trapezium,
$\theta^1$ Ori (HR 1893 -- 6), which consists (somewhat surprisingly) of five
bright components arranged roughly on the boundary of an ellipse
with axes $23\asec\times 13\asec$: see the adaptive-optics mosaic
of Simon et al (1999; their Fig. 1). The components are distributed 
non-uniformly around
this ring, with angular separations (as seen from the center of the
ellipse) of about $90\deg, 30\deg, 30\deg, 120\deg$ and $90\deg$, 
starting from the brightest component at the South and proceeding 
anti-clockwise. But perhaps we should add to these five components
the two main components of $\theta^2$~Ori (HR~1897), about $50\asec$ apart
and about $150\asec$ from the ellipse towards the Southeast. Further,
these 7 components have close sub-components (12, according to 
Preibisch et al. 1999). One component, BM~Ori (HR~1894), is in fact in a
non-hierarchical quintuple {\it sub}-system: BM~Ori iself, which is
an eclipsing and spectroscopic binary, and two further components
making a roughly equilateral triangle about $1\asec$ on the side,
one of which is a very close pair separated by $\ts 0.1\asec$.

\par However, we ought not to ignore several thousand other stars
which are heavily concentrated towards the Trapezium, and evidently
physically associated at some level. Several are within the elliptical
curve (at least in projection), and not just outside it. We choose
to consider the Orion Nebula Cluster (ONC) as just two independent
bright systems, $\theta^1$~Ori~C (HR~1895) and $\theta^2$~Ori~A 
(HR~1897), which are
the only two among the 7 major components which qualify with $\vh\tle 6$.
They are both hierarchical systems of multiplicity 3,
if we ignore their possible relationship to moderately close companions.

\par We identify 17 non-hierarchical systems in our sample. 
Several of these may be simply the three or more brightest stars in a 
cluster; and there are a further few possible binaries that may actually
be just the two brightest stars in a rather distant cluster ($\ts 1\thn$kpc).
Some hierarchical systems appear as non-hierarchical configurations
simply because of projection.

\par Among the bright-star sample are several looser groupings, such
as `OB associations' and `moving groups'. These appear to be the 
last stages in the evaporation of stars from dense star-forming
regions (SFRs) into the general Galactic field. They are probably not 
bound, and so we treat the members of these groupings as individual 
systems, but possibly multiple on a small scale, and if so then usually 
hierarchical.

\section{Procedure}
We searched a number of catalogs, of which we have already mentioned
three (the BSC, MSC and HIP). The BSC is particularly helpful in view of
its copious notes. We also used the 8th spectroscopic-binary
catalog (BFM; Batten, Fletcher \& McCarthy 1989) and the 9th (SB9;
Pourbaix et al 2004). These two catalogs are also very helpful in view of 
their copious
notes. We added the CCDM catalog (Dommanget \& Nys 2002) of close multiples
(many of which however are optical rather than physical multiples)
and the 6th catalog  (VB6) of visual-binary orbits (Hartkopf et al 2000). 
We also incorporated information from the GCVS
(Samus et al. 2004) on eclipsers and ellipsoidal variables, and from
the CHARA survey of speckle observations (Hartkopf et al. 1989, 2000).
\par The Hipparcos experiment also generated a catalog of double and
multiple star measurements, HDMC. It contains mostly the known visual
binaries, but also new systems discovered by Hipparcos.  HDMC frequently 
obtained solutions for
pairs that gave the same parallax and proper motions for close pairs,
thus normally confirming their physical association.

\par We included two Tables of data on astrometric binaries by Makarov 
\& Kaplan (2005; MK), where they considered systems whose Tycho and 
Hipparcos parallaxes and proper motions showed nonlinear behavior with 
time, i.e. acceleration, suggesting astrometric binaries. There are 348 
systems containing such `astrometric accelerators' in our catalog. We also  
included 
Tables from the Galactic Kinematics catalog of Famaey et al. (2005; GKC),
who considered the space motions of a large number of HIP targets and
noted those for which there was evidence of radial-velocity variations
of an orbital character. Several have published orbits, but many more 
are from the private collection
of R. F. Griffin. For several systems with composite spectra, consisting
typically of a red giant and an A or late B star, we give the spectral
types obtained by R. E. M. Griffin (soon to be published) as a result of
careful disentangling of the two spectra; these types are often rather
different from those normally quoted.

\par In addition to GKC above, we used several catalogs of radial velocites 
(Andersen \& Nordstr{\"o}m 1983, Nordstr{\"o}m \& Andersen 1983, 1985, 
Grenier et al. 1999a,b) for B, A/F and G/K/M stars in the
Southern sky. They used the results of several radial-velocity
measurements per star to determine significant variation between
observations. Although there were too few observations to establish
orbits, they found several significant variables, identified as
`VAR'. They also found several marginal cases, indicated as `VAR?'.
We included the former as binaries (or sub-binaries), but left out
the latter. We included the catalogs of Duquennoy \& Mayor (1991)
who considered many multiple systems containing bright solar-type
stars; of Harmanec (2001) for binaries including Be stars; of Aerts
\& Harmanec (2004) for binaries including pulsating stars; of
Parsons (2004) for triple systems with cool giants and hot dwarfs;
and of Lindroos (1985), who attempted to distinguish between physical
and optical multiples among many bright and wide systems.
  
\par Another catalog was a private one maintained by one of us (PPE), 
with about 3000 entries taken from the literature in the interval 1975 
-- 2005; this catalog concentrated on systems which contained stars 
evolved beyond (or not yet up to) the main
sequence, but also contained many bright systems of main-sequence stars.
We do not reference this catalog directly, but instead refer
directly to those papers in it which supplied data that was different
from (and, as we judged, better than) data from the principal catalogs
mentioned above.

\par Following McClure (1983) and Boffin \& Jorissen (1988), we assume
that {\it all} Ba stars are binaries, with a white-dwarf component.
Many Ba stars have indeed been determined to have spectroscopic
orbits, and in a small number a white dwarf has actually been detected
in the UV. It is surprising to us, however, that among the 52 Ba stars
in our bright sample only 10 have known orbits. Three of these have known 
WD components: $\xi^1$ Cet (HR 649), $\zeta$ Cyg (HR 8115) and 
$\zeta$ Cap (HR 8204). However the case for binarity is not just that 
some are confirmed binaries, but much more strongly 
that a physical mechanism exists to explain the Ba anomaly in terms of
binarity, specifically with a white-dwarf companion, and that the anomaly 
is very hard to explain otherwise.

\par The situation is rather different for S stars. Some of these may
be evolved Ba stars, and so also with white-dwarf companions, whether
seen or not. But others could be `intrinsic' S stars, having produced
their own abundance anomalies internally. We find only 4 S stars in our
bright sample (including one with spectrum M4IIIS); two of these have 
known WD companions. The other two have visual companions, which are 
however too far away to be likely WD remnants
of s-process donors. Perhaps the systems are triple, but since they
might instead contain intrinsic S stars we consider them for the present
to be merely binary.

\par For pairs of stars that are fairly close but might be optical rather
than physical, we define a propinquity parameter 
$$X \equiv \log\rho + 0.3V-3.95\z,\eqno(1)$$
where $\rho$ is the separation in arcsecs and $V$ is the magnitude of the
fainter component. We expect $X\tls 0$ if the component is sufficiently 
bright and sufficiently near that only $\ts 1$ such coincidence is likely 
in 5000 cases. The model is based on the
approximation that the number of stars brighter than $V$ is
$$ \log n \ts 3.7 + 0.6(V-6.0)\ ,\eqno(2)$$ 
and that they are randomly distributed over the sky. Many pairs satisfy the 
propinquity test, and most of these are already fairly well established as 
either having measured visual orbits, or common proper motions. However, a
handful are equivocal: they may satisfy the propinquity test by a considerable
margin, and yet the proper motions from HIP or HDMC, and even the parallaxes,
may be quite widely different. Often, though different, they are not
significantly different because the measurement errors happen to be several
hundred times larger than is normal; it is not clear why the errors are so
much larger in a few cases. We tentatively identify a handful of systems
where we suspect that the multiplicity is 3 rather than 2, and that this
is responsible for (a) discrepant proper motions, and (b) unusually
large errors.
\par On the other hand, faint physical components with $X\tg 1$ are
effectively lost in the stellar background, especially near the Galactic plane.
Without further work on proper motions, e.g. L{\'e}pine \& Bongiorno (2007), 
the propinquity test alone biases the observed statistics towards bright 
companions.
\par We accept that some of the systems with negative $X$ are 
optical, one (HR~2764) despite having $X\ts -0.7$. The parallaxes in this 
case are very different, but are roughly in agreement with the 
parallaxes expected from spectral types.
The spectra are K3Ib-II and F0V, and yet the magnitudes differ by only
1.8. The parallaxes differ by a factor of 13, which is no doubt very
uncertain since one is $0.001\asec$; but it seems more reasonable that the
luminosities differ by a factor of $100 - 1000$ than merely 5. A further
optical pair (HR 6008/9) has $X$ substantially less than zero; two exceptions
in $\ts 5000$ are acceptable statistically, just.
\par Parsons (2004) has identified a substantial number of $n\tge 3$
systems among stars whose spectral energy distribution (SED) has been
measured over a wide wavelength range ($0.13 - 0.9\mu$), putting together
data from IUE, Tycho, and ground-based measures. The presence of two (at
least) separate sources, one hot and one cool, is much more evident with 
such a range than in an SED that is limited to the classic UBV range
($0.39 - 0.6\mu$). It is often possible to determine the two temperatures
separately, by fitting the SED to judicious combinations of single-star
SEDs, and this also determines the relative luminosities and radii. Since
the parallaxes are known from Tycho/Hipparcos as well, the absolute
luminosities and radii are known, and can be compared to theoretical
isochrones. Parsons finds several systems where the hot star is too bright,
compared with a theoretical model, by a factor of $\ts 2$, and concludes 
that the hot component is a sub-binary of two roughly equal components.
Parsons (2004) lists 19 systems which feature in our catalog. In
four of them the hot component is already known to be a sub-binary of short 
period, either spectroscopically or photometrically; although in one of these 
four ($\beta$ Cap, HR 7776) the sub-binary is single-lined and therefore 
the unseen sub-component cannot be contributing significantly to the SED.
However HR 7776 (Table 1) appears to be sextuple, and it is possible 
that the nearest of the three remaining components, at $\ts 0.8\asec$,
contributes something.
\par In estimating the multiplicity for a particular system in the present
compilation, we use a different, and weaker, criterion than the MSC for
the certainty with which the multiplicity has been determined so far: in legal
terms, our criterion is roughly `on the balance of probabilities' rather
than `beyond reasonable doubt'. Given our present knowledge, we ask what is
the most probable multiplicity. Clearly in some marginal cases the multiplicity
with the highest probability may have a probability only slightly over 50\%.
For inclusion in the MSC, the criterion was normally stronger: that radial
velocity orbits had actually been determined, for example, rather than
suspected on the grounds of significant radial-velocity variability.

\section{Results}

Table  1 illustrates  our  results.  The  main  body of  the Table  is
available  electronically, and  in the  printed version  here  Table 1
includes only  a few examples,  with a range of  multiplicities.  Many
systems consist  of two  or more  HR entries; we  list them  under the
largest relevant  HR number (Col.~1).  In a  Table of cross-referenced
names (not  shown here but  available electronically) we give  the HR,
HD,  HIP  and  other   identifiers  for  all  components that have
HR numbers, so multiple systems which include several HR numbers can be
easily found.   For systems at  the end, which  have no HR  number, we
give a `pseudo-HR' number greater than 9200 and prefixed by P. These
are identified with genuine names (HD, HIP, \dots) in the same
cross-reference table. 
\par Col.~2
of  Table~1  lists  what  we   consider  to  be  the  most  reasonable
multiplicity.   A multiplicity  $n$ that  is $\tge  3$, and  {\it not}
followed by a query or colon, is the same as in the MSC (316 cases); a
colon indicates that  it is in the MSC with a  different $n\tge 3$ (13
cases),  and  a  query  indicates  that  it is  not  in  the  MSC  (75
cases). Most of  the last category are systems where  we feel that the
balance of probability lies with $n\tge 3$ but that the information is
not so compelling as to make the multiplicity say 95\% certain.

\par Col.~3  contains some reference  letters, defined in  more detail
shortly.  Col.~4  contains the parallax from Hipparcos,  except that a
Hipparcos  parallax $\tl  0.001$, including  values like  $-0.002$, is
replaced by  0.001 automatically.  Col.~5 contains  our description of
the configuration  of the system  using nested parentheses  in roughly
the format suggested by Evans (1977). For each individual component we
give a magnitude and a spectral  type, where we can find them, and for
each pair  of components we  give either a  period or a  separation in
arcseconds, where  we can  find them. Our  reason for  preferring this
notation  here is  that, from  experience, it  can summarise  a system
sufficiently clearly that one can  readily see where each component is
in the hierarchy, and yet it only takes one line per system. To convey
substantially more information, both  about the sub-components and the
sub-orbits, we would need at least one line for each sub-system, as in
the MSC.
\par Although the one-line summaries exemplified in Table 1 may not look
machine-readable, they are. A short code along with the data allowed
the following Tables (2, 3 and 4) to be generated automatically. Various
subsets, e.g. systems with white dwarfs, or with semidetached sub-binaries
can be quickly identified.

\par The magnitudes come from HIP if available, and the spectral types 
generally from the BSC or the MSC if available. HIP magnitudes are given to 
2 decimal places (d.p.) and other magnitudes, mostly Johnson V, to only one. 
Sometimes combined magnitudes of sub-systems are given as well, and always the 
overall combined magnitude. Sometimes we have to combine Johnson magnitudes
with HIP magnitudes to reach a combined magnitude; but the Johnson
magnitudes are usually quite faint and so we treat the combination
as a HIP magnitude.

\par The separation is given to 3 d.p. if it 
comes from Hipparcos, and only 2, 1 or 0 if from another source. The 
period is in days if it comes from a spectroscopic or eclipsing binary; 
the reference is to SB9 or GCVS by default, but the letter `s' implies
an individual reference listed in the electronic version. The period
is in years if it comes from a visual binary; the reference is to
VB6 by default. 
Most systems with multiplicity $n\tge 3$ come from the MSC, by default, 
but the letter `s' implies an individual reference, and in some cases 
letters such as AE (see below) explain why we think it is triple, though not
with as high probability as would warrant inclusion in the MSC.

\par  Some published
visual orbits have periods in excess of 300 yrs. We characterise all
of these by a separation rather than a period, feeling that an orbit
should be seen revolving at least once (and preferably twice) before being 
considered reliably determined. In fact several visual orbits with
periods substantially less that 300 yrs are quite tentative.
\begintable*{1}
\caption{{\bf Table 1.} Sample Configurations of Bright Multiple Systems.}
\halign{%
\rm#\hfil&\rm#\hfil&\rm#\hfil&\rm#\hfil&\rm#\hfil\cr
mHR&$n$&ref.&plx&configuration\cr
\noalign{\vskip 10pt}
     4:&  1 &     &0.009 &5.71G5III\cr
    91:&  3 &s    &0.002 &5.55(5.95(B5IV + ?; 25.42d, e=.12) + 6.84; 152.7y, e=.10)\cr
   120:&  2?&A    &0.022 &5.75(F2V + ?; ?)\cr
   136:&  6 &     &0.021 &3.42(3.68(4.33(B9V + 13.5; 2.4$\asec$) + 4.55(A2V + A7V; 44.66y, e=.74); 27.060$\asec$) + 5.09(A0V + A0V; .1$\asec$); 540$\asec$)\cr
   142:&  3 &     &0.048 &5.32(5.61(F8V + ?; 2.082d) + 6.90G0V; 6.890y, e=.76)\cr
   152:&  2 &R    &0.005 &5.26(K5III + ?; 576.2d, e=.30)\cr
   165:&  3 &R A  &0.032 &3.43((K3III + ?; 20158d, e=.34) + 13.0M2; 28.7$\asec$)\cr
   233:&  3?&G    &0.004 &5.47(G8IIIa + (B9V + ?; 1.916d); 2091d, e=.53)\cr  
   382:&  2?&b    &0.001 &4.95(5.11F0Ia + 7.08B6Ib; 134.061$\asec$)\cr 
   439:&  2 &C    &0.002 &5.82(K0Ib + B9V; .11$\asec$)\cr
   553:&  2 &s    &0.055 &2.70(A5V + G0:; 107.0d, e=.89)   \cr
   629:&  2?&h    &0.012 &5.67(6.07B9V + 6.95A1Vn; 16.690$\asec$)\cr
   958:&  2 &R G s\ &0.004 &5.68(K0II + A7III; 115.0d)\cr
  1556:&  2 &s    &0.006 &4.74(WDA3 + S3.5/1-; ?) \cr
  1564:&  4 &R R s\ &0.031 &5.28((5.67F0IV + ?; ?) + 6.59(F4V + ?; ?); 12.500$\asec$)  \cr
  1788:&  5?&s E E\ &0.004 &3.29(3.58((B1V + B2e; 7.990d, e=.01) + (B: + ?; 0.864d); 9.50y, e=.2) + 4.89B2V; 1.695$\asec$)   \cr
  2788:&  3 &A E s\ &0.023 &5.79((F1V + G8IV-V; 1.136d, SD) + ?; 93.89y, e=.50)\cr                                         
  3963:&  3?&h R s\ &0.007 &5.91(6.30(B8V + ?; ?) + 7.24B9III-IV; 21.200$\asec$) \cr              
  4621:&  5:&A A b\ &0.008 &2.32(2.52(B2IVne + ?; ?) + 4.40(B6IIIe + ?; ?) $@$ 325$\deg$, 267$\asec$ + 6.5B9 $@$ 227$\deg$, 220$\asec$)\cr
  4908:&  2?&L    &0.002 &5.37(O9Ib + 11.8K0III; 29.1$\asec$)\cr
  5340:&  1 &s    &0.089 &0.11K1.5IIIFe-0.5    \cr
  6046:&  2 &R A s\ &0.005 &5.77(5.7K3II + 8.7K0IV-V; 2201d, e=.68)   \cr
  7776:&  6 &P O  &0.009 &3.14(3.21((G8II + 7.2(B8V + ?; 8.68d, e=.36); 1374d, e=0.42) + 8.3; .8$\asec$) + 6.09(6.16A0III + 9.14A1; .68$\asec$); 205.3$\asec$)\cr
  9072:&  1?&s    &0.031 &4.12F4IV                    \cr
 P9203:&  3?&H b  &0.004 &5.79(6.44(B5/6V + ?; ?) + 6.67B8/9V; 129.490$\asec$)      \cr      
 P9207:&  1 &     &0.007 &5.99M8IIIvar  \cr
}
\tabletext{The full Table is to be found at \dots.
\pn For systems containing more than one HR component, we use the maximal HR no., called `mHR' (Col. 1).
If there is no genuine HR no. in the system, we give a `pseudo-HR' no., $\tge 9201$, prefixed by P.
The corresponding HIP and/or HD numbers can then be found in the cross-reference Table 2. One
example shown here, P9203, is HIP 32256 and 32269. We identify only 7 pseudo-HR systems that qualify
for our sample. Col. 2 (n) is the estimated most probable multiplicity.
\pn In col. 3, the letters refer to various sources, as described in the text. The
absence of a reference letter also implies particular catalog sources, as also
described in the text. Col. 4 is the parallax from Hipparcos, but (a) rounded up to $0.001\asec$ if
less than this (including zero and negative values) and (b) averaged, if two or more components
appear to be part of the same system and yet have somewhat different listed parallaxes (eg HR 126/127/136).}
\endtable

\begintable*{2}
\caption{{\bf Table 2.}  Sample from Cross-Reference Table.}
\halign{\rm\hfil#&\rm\hfil#&\rm#\hfil&\rm#\hfil&\rm#\hfil&\rm#\hfil&\rm#\hfil&\rm#\hfil&\rm#\hfil&\rm#\hfil&\rm#\hfil&\rm#\hfil&\rm#\hfil&\rm#\hfil\cr
\noalign{\vskip 10pt}
 glo\ \ & gla\ \ &    mHR\ &   HR  \ &     HD  \ &     HiP \ &    GCVS \ & Bayer  \ &Flam-\   &BD/CoD\        &    ADS  \ &  CCDM \      &   IDS/MSC  \ &   cluster\cr
    \ \ &    \ \ &       \ &       \ &         \ &         \ &         \ &        \ &steed\   &/CpD \         &         \ &  /WDS  \     &            \ &          \cr
306.78\ &$-$54.02\ & HR136 \ & HR126 \ &HD2884   \ & HP2484  \ &         \ &betTuc1 \ &       \ & CP$-$63d50    \ &         \ &            \ &  00270$-$6331\ &\cr
306.78\ &$-$54.01\ & HR136 \ & HR127 \ &HD2885   \ & HP2487  \ &         \ &betTuc2 \ &       \ & CP$-$63d50    \ &         \ &  00315$-$6257\ &  00270$-$6331\ &\cr
120.84\ &  0.14\ & HR130 \ & HR130 \ &HD2905   \ & HP2599  \ &kapCas   \ &kapCas  \ &  15Cas\ & BD+62d102   \ &         \ &            \ &            \ &  CasOB14\cr
116.94\ &$-$42.36\ & HR131 \ & HR131 \ &HD2910   \ & HP2568  \ &         \ &        \ &  52Psc\ & BD+19d79    \ &ADS452   \ &            \ &            \ &\cr
114.55\ &$-$55.61\ & HR132 \ & HR132 \ &HD2913   \ &CHP2548  \ &NSV15113 \ &        \ &  51Psc\ & BD+06d64    \ &ADS449   \ &  00324+0657\ &  00272+0624\ &\cr
306.54\ &$-$53.97\ & HR136 \ & HR136 \ &HD3003   \ & HP2578  \ &         \ &betTuc3 \ &       \ & CP$-$63d52    \ &         \ &            \ &  00270$-$6331\ &\cr
 54.20\ &$-$39.26\ &PHR9207\ &PHR9207\ &HD207076 \ & HP107516\ &EPAqr    \ &        \ &       \ & BD$-$02d5631  \ &         \ &            \ &            \ &\cr
206.01\ & 32.32\ &PHR9209\ &PHR9209\ &HD73598  \ & HP42497 \ &NSV4171  \ &        \ &       \ & BD+20d2150  \ &ADS6915  \ &  08399+1933\ &            \ &  NGC2632\cr
206.00\ & 32.34\ &PHR9209\ &PHR9209\ &HD73618  \ &         \ &         \ &        \ &       \ & BD+20d2152  \ &ADS6915  \ &  08399+1933\ &            \ &  NGC2632\cr
206.02\ & 32.34\ &PHR9209\ &PHR9209\ &HD73619  \ &         \ &NSV4174  \ &        \ &       \ & BD+20d2153  \ &ADS6915  \ &  08399+1933\ &            \ &  NGC2632\cr
}
\tabletext{The full Table is to be found at \dots.
\pn The first two columns are galactic latitude and longitude. The next two columns, mHR and HR, give the system number and the component
number, which may differ if a system contains more than one component with an HR number -- for example HR 126/127/136. In a small number
of cases at the end, systems and components have no HR number, and so are given a `pseudo-HR' number, of no significance except to
cross-reference Tables 1, 2 and 3. The remaining columns are standard identifiers, except that Greek letters in Bayer names
are rendered by the first three letters (or if necessary only two) of the usual spelling in English.}
\endtable

\begintable*{3}
\caption{{\bf Table 3.} Sample from Reference Table}
\halign{%
\rm\hfil#&\rm#\hfil&\rm#\hfil&\rm#\hfil\cr
\noalign{\vskip 10pt}
    91\ \ \ &       ApJ655,473&\cr
   553\ \ \ &       Obs108,228&\cr
958\ \ \ &JAA11,491&&\cr
965\ \ \ & Obs109,180\ \ \ & http://www.eso.org/gen-fac/meetings/ms2005/griffin.pdf&\cr
  1556\ \ \ &       ApJ327,214&\cr
}
\tabletext{The full Table of references is to be found at ...
}
\endtable

\par The reference letters in Col. 3 have the  following significance.
By default, i.e. in the absence of a reference letter, a system with 
$n\tge 3$ comes from the MSC, and with $n\tle 2$ from SB9, USNO, BSC, GCVS
and/or CCDM, supplemented by HIP or HDMC. Systems with orbits from SB9/GCVS 
or VB6 are distinguished by having the period in days or in years, 
respectively. When a reference letter {\it is} given it has the following 
meaning:
\pn  A is for an entry that has been identified as a probable astrometric 
binary by Makarov \& Kaplan (2005); this presumably has a long period, 
and so if the system is already known to have a short period (from SB9, 
GCVS, or another source) we assume the system is triple (at least). 
If this makes it a triple that is {\it not} in the MSC, we write
$n=3?$ rather than $n=3$. Quite often it is in the MSC, however, because
there is convincing additional evidence.
\pn  b is for an entry of two or more stars that are (usually) intrinsically 
bright, 
distant, and not very close together. They are close enough together
that their juxtaposition is unlikely to be chance, yet far enough apart
that one can question whether they are in a long-lived orbit. It ends
up as largely a subjective matter whether one thinks these should be listed
as one system, or two or more systems. In some cases they may be
members of a cluster, and (arguably) not much closer together than cluster
stars usually are. We list 8 systems that we qualify with b.
\pn  C is for an entry that comes from CHARA. 
\pn  E is for an entry identified in
the GCVS as having an eclipsing, or eclipse-related, light curve. If
a system merits both A and E we regard it as a probable triple, since
an orbit that gives eclipses will usually be too small to also give
an astrometric indication. Most but not all systems flagged E from the
GCVS have known periods, coinciding with SB9 periods, but those without 
periods are presumably rather uncertain.
\pn  G is for a system whose parameters are taken from the catalog of
R. E. M. Griffin (to be published). These are mostly G/K giants plus
B/A dwarfs; but in a handful of cases she concludes that the dwarf is
itself in a sub-binary.
\pn  H is for a system that we identify, through slightly discrepant proper
motions, as having an extra, unseen, component making the system at least 
triple. There are 6 such
systems; we assign their multiplicities as 3?, indicating that we think that
the probability of the extra body is over 50\%, though nothing like certain
enough for inclusion in the MSC.
\pn  h is for a binary system where we suspect that large but uncertain
discrepancies in the HIP values of parallax and/or proper motion may be
hinting very weakly at a third body. We assign multiplicity 2?, and we identify
25 such systems.
\pn  L is for a system where we have relied on the discussion by Lindroos
(1985) as to whether certain companions, usually rather distant, are 
optical or physical, and arguably pre-Main-Sequence if physical. 
\pn  P is for a system noted by Parsons (2004) as containing a hot
component that may be two roughly equal components.
\pn  R is for an entry identified as a probable or certain spectroscopic
binary by Famaey et al. (2005), Andersen \& Nordstr{\"o}m (1983)
Nordstr{\"o}m \& Andersen (1983, 1985) or Grenier et al (1999a,b).
\pn  s is for a specific reference which is given in the electronic 
version of the Table.

\par Although nested parentheses are a good way of displaying 
hierarchical character, they do not work for non-hierarchical systems. 
For these we use the following notation. Suppose that $X, Y$ and $Z$ 
are a non-hierarchical triple. Then we write it as
$$    X\  +\  Y @ \ \theta,\  \rho\  +  \ Z\ @\  \theta^{\thn\prime}, 
\ \rho^{\prime}$$
with position angle $\theta$ in degrees and separation $\rho$ in
arcseconds. Both are measured from component $X$. An example in the
printed Table 1 is HR 4621.

\par We attach considerable weight to Hipparcos observations, but we should 
note that Hipparcos claims that $\alpha$ Boo (HR 5340), at a distance of only
$11\thn$pc, is a binary with $\Delta \vh =3.33$ and separation $0.260\asec$. 
Griffin (1998) has shown that this is very difficult to reconcile with
the record of radial-velocity measurements, which are many and accurate.
It is hardly conceivable that such a companion would not have produced a
recognisable  radial-velocity curve over the last 50 years. Nevertheless
the issue may still be open: see S{\"o}derhjelm \& Mignard (1998) 
and Verhoelst et al. (2005). We list it as single, believing that the
companion is an artefact.
\par We discuss the small subset of entries listed in Table 1, mainly to 
aid the reader in seeing why we put forward the multiplicities listed 
there. We refer to the electronic version of the Table as the `EV'.
\sep HR 4: The absence of reference letters implies that the parallax and
magnitude come from~HIP, and the spectral type from BSC; and there is no
clear, or even tentative, indication of any companion in any of the 
catalogues consulted, or indeed in any paper we have seen.
\sep HR 91: An absence of reference letters, coupled with the fact that
the multiplicity is $n\tge 3$, would mean that the basic source of the data is
the MSC; but the magnitudes come from HIP. However the reference letter
`s' means that some data (the shorter orbit) comes from a paper referenced
in the EV, and in fact the longer period, taken from VB6, is moderately
different from the MSC.
\sep HR 120: This is an `astrometric accelerator' from MK, which we 
therefore suppose to be a binary with an unseen companion and unknown 
(but presumably fairly long-period) orbit.
\sep HR 136: This is actually 3 HR stars (126/127/136), but we label it by
the largest HR number. The system is a wide but apparently hierarchical
triple of three close binaries. The three HIP parallaxes differ by a surprising
20\%, and the three proper motions are somewhat different, though not
alarmingly so. Probably the internal motions of the three binaries account
for most or all of this. The system is accepted as real by the MSC. 
Our parallax is the mean of the three.
\sep HR 142: With no reference letter (unlike HR 91) the data is straight
from the MSC, except for the magnitude from Hipparcos.
\sep HR 152: This is a radial-velocity variable flagged by GKC; in fact it has 
a known orbit, in SB9 (by default), which we presume is responsible for the 
variation.
\sep HR 165: This is both a radial-velocity variable from GKC, and an
astrometric accelerator from MK. We assume that both of these are accounted
for by the spectroscopic orbit which is (by default) in SB9.  The separation 
of the third component, quoted to only 1 d.p., comes from the BSC.
However, if the separation were quoted to 3 d.p. it would be
a HIP measurement, where there is normally supporting evidence from common,
or nearly common, proper motion. The unseen companion in the spectroscopic
orbit might in principle be a white dwarf, but the mass function is sufficiently
small that an M dwarf cannot be ruled out.
\sep HR 233: This $n\tge3$ system is not in the MSC, hence `3?'. R. E. M. 
Griffin finds an inner orbit of 1.916$\thn$d. `G' is a reference to R. E. M. 
Griffin (to be published).
\sep HR 382: A pair of intrinsically bright, distant stars, part of the cluster
NGC 457. They are the two brightest members by $\ts 2\thn$mag, and some way
to one side of the cluster center as defined by about half-a-dozen stars of
magnitude 9 -- 10. Their projected separation is $\ts 0.67\thn$pc, about 
$2.5\thn$pc
from the center. They seem to be on the rather broad margin between
indisputable binaries, and pairs or small groups of stars that clearly have
a common origin but may no longer be bound. We note eight such systems,
referenced with letter b.
\sep HR 439: The separation comes from CHARA. Since it is small, it may well
be changing quite rapidly; we simply list the value quoted at the epoch of
measurement.
\sep HR 553: Although the orbit is in SB9, the spectral type of the
companion is not; the `s' means that a specific reference is given for this.
\sep HR 629: The `h' implies that this is a system where the difference in HIP
proper motions is rather larger than we would expect as a result of the
well-known companion at 17$\asec$. However, one of the proper motions is
unusually uncertain. We therefore entertain the possibility that a third body 
is present to disturb one of the two known bodies, but do not feel that
the probability is high enough to raise the multiplicity from 2? to 3?.
\sep HR 958: This is a radial-velocity variable from the GKC, but an orbit is
known which presumably accounts for this. It is not in the SB9 so the flag 
`s' implies a specific reference to it (in the EV). The spectral types are 
referenced to `G' (i.e. R. E. M. Griffin, to be published); they are more 
precise than the  K1IIIep + A6V given in the BSC.
\sep HR 1556: The `s' implies a specific reference (in the EV), to the fact
that the companion is known to be a white dwarf.
\sep HR 1564: This visual binary contains two radial velocity variables,
as indicated by the MSC (as well as by the double entry, implied by RR, in
the GKC).
\sep HR 1788: This complex system is based on a specific reference (De May et 
al. 1996), as implied by `s'. In addition to the
eclipse period of the first sub-binary listed, there is a second period
($0.864\thn$d) which comes from some other component. It is most probably
ellipsoidal variation, and while it might come from the B2V component, the 
second-last component listed, De May et al. conclude that it is probably 
from the component we have indicated.
\sep HR 2788: This is a well-known Algol (R CMa), but it is also an astrometric
accelerator (MK). The latter can hardly be due to the small orbit of the 
Algol, and so we could infer a distant third body. In fact such a body is known,
and listed. There are other systems where
both `A' and `E' make it clear why we suppose them to be triple. There are 
four such AE triples in our catalog. SD in the description stands for
`semidetached'; we note 18 such systems in the entire sample. We flag contact
systems with CT, and there are three such systems.
\sep HR 3963: Hipparcos gives very different parallaxes ($.007\asec$, 
$.001\asec$) 
and fairly different proper motions; thus the system appears to be optical.
Yet a factor of 7 in distance seems entirely at odds with the fact that the
spectral types and magnitudes are fairly consistent with their being at the 
{\it same}
distance. Also they are {\it very} close for an optical pair; we would only
expect one such pair in a sample ten times larger. We suggest that the smaller
parallax is wrong because of an unseen third body, and that this accounts for
the difference in proper motion also.
\sep HR 4621: This is a non-hierachical triple in the MSC, but two components
are astrometric accelerators. Thus we conclude that it may have $n=5$; the triple
is too wide to account for the acceleration without other bodies. 
\sep HR 4908: This apparently improbable juxtaposition of an O supergiant
with a faint K giant companion is suggested by Lindroos (1985) to be a physical
system in which the secondary is still contracting to the ZAMS. Lindroos
suggests several other pre-MS companions to young massive stars.
\sep HR 5340: We have already discussed the fact that this star ($\alpha$ Boo) 
is perceived as binary by HIP, but can hardly be binary in the face of much
radial-velocity information to the contrary. The `s' flag indicates that we
give specific references.
\sep HR 6046: This is both an astrometric and a radial-velocity variable, but
both are presumably accounted for by the known orbit. SB9 gives an 
approximation to the orbit, but our flag `s' points to a specific reference
(in the EV) which we believe is better.
\sep HR 7776: This sextuple is rather complex, with conflicting 
interpretations. We accept that of the MSC. The flag `P' indicates that 
Parsons (2004) considered that the system needs an extra hot component.
\sep HR 9072: The flag `s' points to two specific references which flatly
contradict each other. One says the system is triple, the other single (or at
least not demonstrably multiple). It is probably a $\gamma$ Dor variable,
with intrinsic pulsation capable of masking, or being mistaken for, orbital
motion.
\sep P9203: This is a pair of non-HR stars (HIP 32256 and 32269) which are
rather far apart on the sky and yet which are both sufficiently bright that
the propinquity factor ($X=0.2$) is uncomfortably small for an optical 
pair. Their parallaxes are the same, but their proper motions differ
modestly, to the extent that we invoke an unseen third body to account for
the discrepancy. Their luminosities are uncomfortably equal considering
the difference in spectral types, but not unreasonably equal. If they are
a real pair, their combined magnitude puts them in our catalog, and since
they would then be an $n\tge3$ system we flag them with `n=3?', i.e. not in
the MSC. We assign a `pseudo-HR' number ($\tge 9201$), and give the actual
ID in the electronic cross-reference Table.
\sep P9207: This is a Mira variable whose average magnitude as determined by HIP 
puts it in our sample. It is not in the HR catalog, but is HIP 107516 as given
in the cross-reference Table.

\par Table 4 gives the numbers of systems observed with multiplicity
from 1 to 7, for four samples. The first two are $\vh\tle 6$ and 
$\vh\tle 4$; the last two are the first sample divided into Northern
and Southern hemispheres. Note that the Sun is included as a bright
star in the first two samples, but is not included in either the
North or the South sample.

\begintable*{4}
\caption{{\bf Table 4.} Multiplicity Frequency in Four Samples.}
\halign{%
\rm#\hfil&\qquad\rm#\hfil&\qquad\rm\hfil#&\qquad\rm\hfil
#&\qquad\rm\hfil#&\qquad\rm\hfil#&\qquad\rm#\hfil
&\qquad\rm\hfil#&\qquad\rm#\hfil&\qquad\hfil\rm#\cr
Sample&total&$n=1$&2&3&4&5&6&7&av.\cr
\noalign{\vskip 10pt}
$\vh\tle 6$& 4559& 2718& 1437&  285&   86&   20&   11&    2&  1.53\cr
$\vh\tle 4$&  478&  213&  179&   54&   19&    8&    5&    0&  1.84\cr
North      & 2141& 1233&  697&  140&   52&   11&    7&    1&  1.57\cr   
South      & 2417& 1484&  740&  145&   34&    9&    4&    1&  1.49\cr
}
\tabletext{The last column is the average multiplicity or `companion
star fraction (CSF)', defined as $\sum_n nN_n/\sum_n N_n$. 
If $N_n\propto a^{-n}$, the average is $a/(a-1)$. The first 
two samples include the Sun, the last two exclude it.}
\endtable

\par Table 5 gives the distribution of orbital periods, in bins of width
1.0 in $\log P\thn$(yrs). For wide systems with only a separation listed,
the period is estimated from the angular separation, parallax, and Kepler's 
law, assuming a mass dependent on spectral type. The distribution is
broken up into separate distributions according to the spectral type
of the main component; type B is divided into early B (B0 to B3.5) and
late B. The distribution of period can be seen to be strongly bimodal at 
early types, becoming weakly bimodal, or unimodal with a very broad 
maximum, at later types. The minimum where it appears is generally in
the bins $0.1 - 10\thn$yrs. 
\par Table 6 gives the fractional multiplicity, broken down by the
spectral type of the leading component. We see that O stars (which
for our purposes include two Wolf-Rayet systems) appear to be
substantially more multiple than later types. However there seems
to be little variation in fractional multiplicity among types B -- G.
It is not surprising that our G/K sample should be little different
from our A/F sample, since most of our G/K sample are evolved giants 
and are essentially the same population as the A/F sample (mostly main 
sequence); but there is little variation even for eB and lB, where the 
relation is less close.
\def\zz{\ \ \ \ }
\begintable*{5}
\caption{{\bf Table 5.} Period Distribution in Systems and Subsystems.}
\halign{\rm#\hfil&\hfil #&\hfil #&\hfil #&\hfil #
&\hfil #&\hfil #&\hfil #&\hfil #&\hfil #&\hfil #&\hfil #&\hfil #&\hfil #&\hfil #\cr
$\log P$(yr)&&-3.0\zz&-2.0\zz&-1.0\zz&0.0\zz&1.0\zz&2.0\zz&3.0\zz&4.0\zz
&5.0\zz&6.0\zz&7.0\zz&8.0\zz&total\cr
sp.&&&&&&&&&&&&&&\cr
  O &   0 &   5 &  11 &   5 &   0 &   4 &   3 &   6 &  12 &   1 &   0 &   0 &   0 &  47 \cr
 eB &   0 &  25 &  41 &  21 &  14 &  21 &  26 &  29 &  33 &  21 &   3 &   0 &   1 & 235 \cr
 lB &   0 &  25 &  53 &  24 &  20 &  43 &  62 &  63 &  48 &  16 &   8 &   0 &   0 & 362 \cr
  A &   0 &  27 &  62 &  25 &  46 &  78 &  78 &  66 &  47 &  24 &   9 &   0 &   0 & 462 \cr
  F &   0 &  14 &  33 &  24 &  39 &  46 &  61 &  44 &  26 &  14 &   3 &   1 &   0 & 305 \cr
  G &   1 &   7 &   9 &  20 &  40 &  38 &  49 &  38 &  32 &  23 &   6 &   1 &   0 & 264 \cr
  K &   0 &   4 &   4 &  12 &  56 &  35 &  35 &  42 &  35 &  26 &   4 &   0 &   0 & 253 \cr
  M &   1 &   0 &   0 &   3 &   7 &   6 &   4 &   9 &   9 &   3 &   3 &   0 &   0 &  45 \cr
sum &   2 & 107 & 213 & 134 & 222 & 271 & 318 & 297 & 242 & 128 &  36 &   2 &   1 &1973 \cr
}
\tabletext{The first column gives the spectral type of the dominant body 
in the system: eB means early B, i.e. B0 -- B3.5, and lB means later B.
Wolf-Rayet stars (2) are included under O; S and C stars are included under M.
The first column of integers gives the number of systems and sub-systems with 
$\log P$(yr)$\tle -3.0$; the second for
$-3.0\tl\log P\tle-2.0$, etc. The two shortest periods are a contact binary 
subsystem of the G dwarf 44 Boo (HR 5618), and a cataclysmic binary subsystem 
of the M giant CQ Dra  (HR 4765). Long periods are estimates from the
angular separation, distance, and Kepler's law.}
\endtable
\begintable*{6}
\caption{{\bf Table 6.} Fractional Multiplicity by Spectral Type.}
\halign{\rm#\hfil&\hfil #&\hfil #&\hfil #&\hfil #
&\hfil #&\hfil #&\hfil #&\hfil #&\hfil #&\hfil #\cr
sp.&1&2&3&4&5&6&7&av.&tot$_6$&tot$_4$\cr
 O&\zz 0.342&\zz 0.263&\zz 0.184&\zz 0.105&\zz 0.053&\zz 0.053&
\zz 0.000&\zz 2.42&\zz\zz38&\zz\zz 7\cr
eB& 0.531& 0.312& 0.115& 0.020& 0.015& 0.002& 0.005&  1.70&  401&   84\cr
lB& 0.541& 0.329& 0.093& 0.032& 0.002& 0.003& 0.000&  1.64&  653&   63\cr
 A& 0.540& 0.362& 0.068& 0.018& 0.006& 0.005& 0.000&  1.61&  928&   90\cr
 F& 0.526& 0.379& 0.062& 0.026& 0.007& 0.000& 0.000&  1.61&  578&   49\cr
 G& 0.550& 0.370& 0.056& 0.020& 0.002& 0.002& 0.000&  1.56&  588&   60\cr
 K& 0.709& 0.253& 0.031& 0.008& 0.000& 0.000& 0.000&  1.34& 1043&   99\cr
 M& 0.821& 0.155& 0.021& 0.003& 0.000& 0.000& 0.000&  1.21&  330&   26\cr
}
\tabletext{Early B stars (B0 -- B3.5) are called `eB'; later B stars are 
called `lB'. Wolf-Rayet stars are included under O; S and C stars under M. 
The third last column is
the average multiplicity; the last two columns are the total number, in the 
larger sample ($\vh\tle 6$) and the smaller sample ($\vh\tle 4$) respectively.}
\endtable
\section{Completeness, Detection Efficiency and Selection Effects}

\par One might suppose that it would not be difficult to identify the complete 
list of systems with $\vh\tle 6.00$, supposing for the moment that we are not 
yet interested in the individual multiplicities. There are however minor 
issues, which make for an uncertainty of perhaps $\pm 10$ in our list of 4559 
systems. The main one is deciding whether two or more stars, each fainter than
$\vh = 6.00$, are a real system or not. We believe that there is no answer to 
this that everyone would accept; and the main reason for this is the 
long-range character of the gravitational force, something that we cannot vary
or work around. Nevertheless, we have tested a number of algorithms in
which brightness is added up for all Hipparcos targets that are closer together
than some limit -- either an angular limit, or a linear limit involving
parallax. With appropriate choice of limit we can pick up clusters
as spread-out as the Hyades, but we have concentrated on smaller limits,
in particular 180$\asec$. This produces 130 pairs or higher multiples,
in addition to many more that qualify only as single targets in HIP but as
multiples in the HDMC catalog. The great majority of the 130 are well-known
`systems' whose reality has been investigated over decades or centuries,
and where there is rather little doubt of the reality or otherwise. But
we always come across a few near the margin, wherever and however we
might try to define the margin. The smallest angular separation that we
accept as optical (among pairs of nearly equal brightness) is $\ts 27\asec$,
for HIP 35210/35213 (HR 2764) and also for  HIP 79043/79045 (HR 6008/6009),
and the two largest that we accept provisionally as real are
$7860\asec$ for $\alpha$ Cen (HR 5460), and $\ts 7030\asec$,
for $\alpha$ PsA (HR 8728). The linear separation  
in $\alpha$ PsA is slightly more than 0.25$\thn$pc, and we can wonder 
whether such a system can survive even for one full rotation. The age of
$\alpha$ PsA is estimated as $200\pm100\thn$Myr by Di Folco et al. (2004).
If the two main bodies have been moving apart steadily over that interval,
i.e. not orbiting but escaping, they have been doing so at the very low 
velocity of $0.0025\thn$km/s, much less than the actual escape velocity. 
On the other hand, if they are orbiting, they have survived rather 
surprisingly for $\tgs 20$ orbits of $\ts 6\thn$Myr each. But despite the
uncertainty of such cases, we feel that only about 10 in (or not in) our 
entire sample are rather marginal either way.

\par Detection efficiency, such as the difficulty of observing long-period 
spectroscopic binaries and short-period  visual binaries, will make the 
distributions of multiplicity, and of periods, differ from the true 
distribution. One of us (PPE) proposes to investigate this further using a 
Monte Carlo procedure that allows one to (a) construct a magnitude-limited
sample of systems with multiplicity 1 to 8 according to some hypothetical
distribution of masses, ages, multiplicity and periods, and (b) to 
`theoretically observe' the members of this sample, making estimates of the 
efficiency of various observational procedures. This can map the original 
multiplicity into at least a lower limit to the observed multiplicity. It is 
difficult to believe that observations to date can constrain the number of T 
dwarf companions that an O star might have, but some less extreme pairings may
be capable of being ruled in (or out).

\par Selection effects often interact with detection efficiency: if a given
technique is known to work well for certain types of star, those types of
star are more likely to be investigated with that technique. Thus excellent 
radial-velocity orbits of quite long period G/K binaries, say 1 - 30$\thn$yr, 
are known, but very few in this period range are known for O or B stars, 
because their broad lines preclude measurement to the necessary accuracy. 
Nevertheless there may be the same proportion of $1-30\thn$yr binaries in 
both subsamples. 
\par While the limits of various observing techniques in detecting binary
companions can be modeled reasonably well, the extent of their application
to our sample remains unexplored. For example, radial velocities of almost
all bright stars have been measured several times, but the accuracy and
time coverage vary to such an extent that we cannot apply uniform criteria
to the sample as a whole.  Many subsystems which are detectable 
spectroscopically still remain undetected. Similarly only a fraction of 
bright stars has been observed interferometrically.

\par We notice in Table~3 that the period distribution is bimodal 
at early spectral types, but unimodal, with a very broad maximum, at later 
types. That might be because the intermediate periods for O stars, 
$0.1 - 100\thn$yr, are too close to recognise visually (although some are 
resolved), and too wide to recognise spectroscopically (although some also have
been recognised). A preliminary version of the Monte Carlo code (Eggleton et al. 2007)
suggested that the apparent bimodality of O-star periods can be explained this 
way; we should bear in mind the result (Table~4) that O stars appear to be more
highly multiple than later stars, so that it can be not improbable for an 
O-star system to have at least three bodies, one pair of which might have a 
period in the `missing' range.

\par There are indications in Table~2, or equivalently Fig. 1, that many 
multiples remain to be detected. On the one hand, the 4-mag-limited sample of
Table 4 shows frequency $N_n$ of multiplicity $n$ dropping substantially less
rapidly (roughly, $N_n\ts 2.3^{-n}$) than the 6-mag-limited sample (roughly, 
$N_n\ts 3.4^{-n}$). We would expect at most a small effect here. The brighter 
sample is on the whole nearer, so that visual multiples are more readily 
recognised, and biased to slightly more massive stars, which tend to be more 
highly multiple. But preliminary attempts at Monte-Carlo modeling (Eggleton et 
al. 2007) suggest that this can only account for perhaps a third of the change 
in average multiplicity as seen between the two samples. We suggest that most 
of the difference is due to the fact that over the last 3 centuries the 
brightest stars have been studied most carefully. Interestingly this may now 
be the opposite of the truth. Modern detectors can be so sensitive that
observers {\it avoid} the brightest stars.

\par The comparison of the Northern and Southern half-samples in Table 4 tells 
a possibly similar story. The fact that the Northern sample has a modestly 
higher average multiplicity than the Southern sample may be because the number 
of telescopes, and of observers, in the Northern hemisphere has always been 
substantially larger. The difference between the Northern and Southern samples 
is not significant at the 99\% level, but it is significant (C. A. Booker
2008, p.c.) at the 95\% level, according to a $\chi^2$ test. The preliminary 
Monte Carlo model of Eggleton et al. (2007) suggested that the true average 
multiplicity would have to be above 2.0 if the observed average multiplicity 
for the complete sample (top line of Table~2) is to be in excess of 1.5.

\par  In  Table 5  we  do not  discriminate  between  inner and  outer
periods,  although  it would  not  be  difficult  in principle  to  do
so. Such discrimination would reveal that  close pairs prefer
to  be inner  components  of higher-multiplicity  systems rather  than
pure $n=2$ binaries. The present catalog, as well as the MSC, contains
rich information on the statistics of periods and mass ratios at
different hierarchical levels. 
\centerline{\psfig{figure=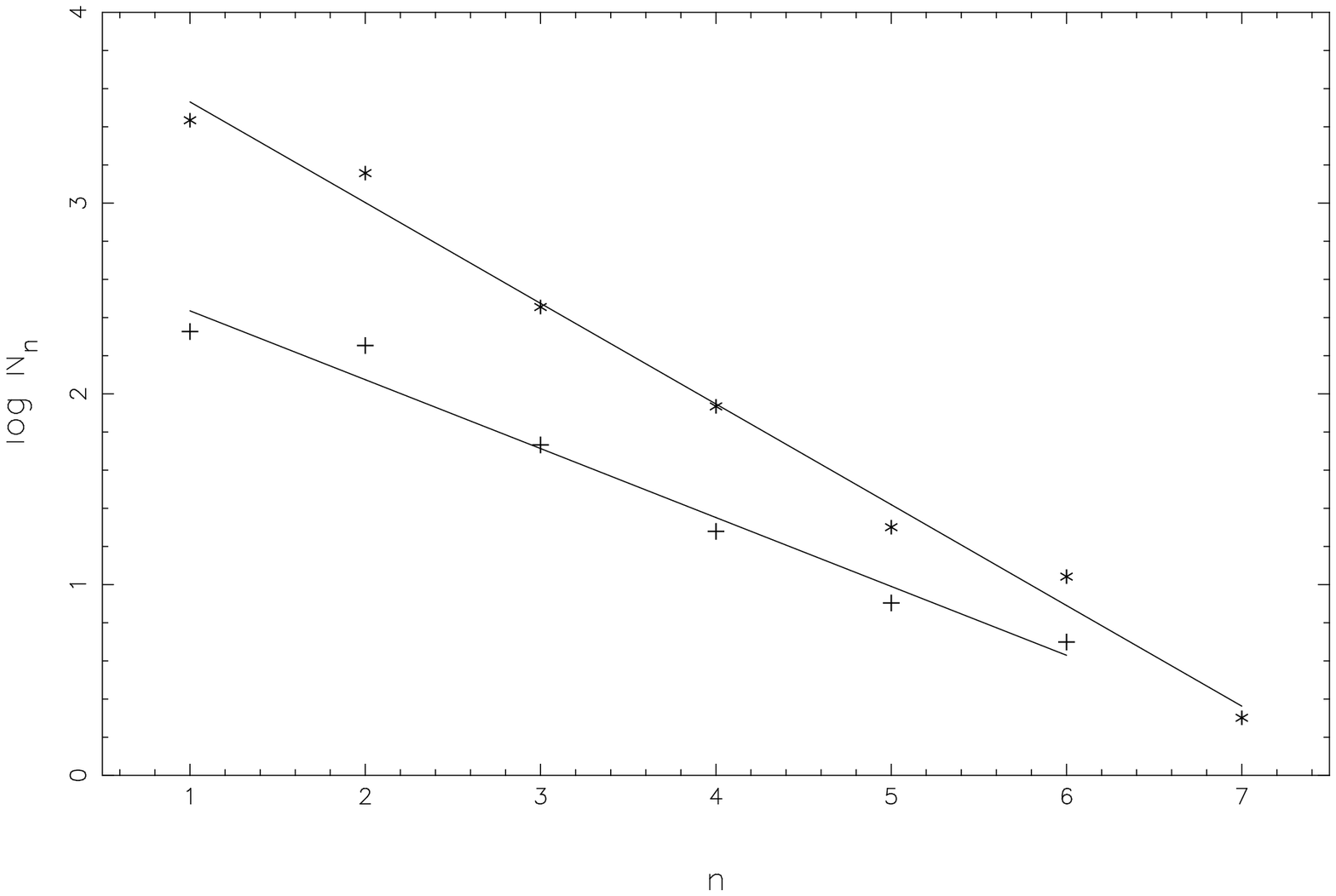,height=2.2in,bbllx=0pt,bblly=390pt,bburx=560pt,bbury=780pt,clip=}}
{\small Fig 1 -- Log frequency as a function of multiplicity, from the
first two lines of Table 1. Least-squares fits are plotted, with slopes
corresponding to -3.37 for the larger sample, and
-2.30 for the smaller sample.}

\section{Discussion}
\par We see three main areas where statistics of the multiplicity and period
distributions for a complete magnitude-limited sample is potentially useful. 
Firstly, they can constrain
modeling of the star-formation process. Star formation lacks any clear
initial conditions, but at least it has some reasonably clear `final
conditions', and models can be tested against their ability to produce
the kind of systems that are actually seen, in the right proportions.
Secondly, the model can serve as the initial condition for subsequent
nuclear evolution; and as we amplify below there is also the possibility 
of dynamical evolution within triples and higher multiples that can
influence their nuclear evolution. Thirdly, we believe that such a model
can also be helpful in computing the evolution of star clusters under
the joint effects of nuclear evolution and N-body gravitational dynamics.
Although the dense environment of those star-forming  regions that
generate clusters, i.e. that make a transition from a gas-dominated entity 
to a star-dominated entity of much the same size,
is likely to prevent the existence of the widest multiples that our
sample contains, it is a reasonable first hypothesis that the shorter-period
multiples will not be very different. If in fact an N-body calculation
were started with statistical multiplicity as in our sample, presumably the
wider systems would be rapidly broken up; but the closer systems could
persist, and the presence of rather close primordial triples and
quadruples would influence significantly the evolution of the cluster.

\par Regarding the second area, Eggleton \& Kiseleva (1996) have enumerated a 
number of ways in which the existence of primordial triples can allow 
evolutionary channels that are different from what can be obtained from only 
binaries. We mention two here. Firstly, triple stars in which the two orbits
are misaligned can be subject to the dynamical effect of Kozai cycles
(Kozai 1962),
and these in turn can allow tidal friction to become important in the
course of $10^6 - 10^9\thn$yrs and cause the inner orbit to become
smaller (Mazeh \& Shaham 1979, Eggleton \& Kiseleva et al. 1998,
Eggleton \& Kisseleva 2006). We call
this process KCTF (Kozai Cycles with Tidal Friction). Tokovinin et al. (2006) 
have noted, using a maximum-likelihood method, that in a sample of 
spectroscopic binaries with periods $\tl 2.9\thn$d, $96\pm 7\%$ have third 
bodies, as compared with $34\pm6\%$ in the period range $13 - 30\thn$d.  
Fabrycky \& Tremaine (2007) have computed the effect of KCTF on a Monte Carlo
ensemble of triples, and shown that indeed it can produce an accumulation
at short inner periods. Pribulla \& Rucinski (2006) have noted that as many as 
42\% of contact binaries appear to be in triples (and arguably 59\% in a more 
thoroughly examined subsample), and it could be that KCTF has contributed
to this; although we probably need the additional influence of magnetic
braking, also with tidal friction (MBTF), to drive fairly close low-mass 
binaries generated by KCTF to contact on a timscale of $\tls 10^9\thn$.
In our primary sample, line 1 of Table 4, there are 95 binaries with
$P\tl 3\thn$d, and 60 of these are in systems with $n\tge 3$; this is
a much higher proportion than for longer periods.
\par The second effect is the production of `anomalous binaries'. In
short-period binaries we can expect that a merger of the two components
is a fairly common event. Case A systems can evolve conservatively only if the
initial mass ratio is fairly mild ($1\tg q\tgs 0.6$; Nelson \& Eggleton 2001),
and if $q$ is not this mild then a merger seems quite a likely event.
It would be hard to determine that a particular currently-single star
is a merged remnant of a former binary, although some Be stars that are
apparently single might be such remnants. But within a primordial triple 
system, it is possible that such a merged remnant would be identifiable,
because the wide binary that remains after the merger of the close
sub-binary would be expected, in at least some circumstances, to show
an anomaly where the two components appear to be of different ages. R. E. M. 
Griffin (to be published) has found a number of such apparently anomalous
systems, of which $\gamma$ Per (HR 915) is an example. Although the
mass ratio, obtained from careful deconvolution of the two spectra
(G8II-III + A2IV; $5330\thn$d, $e=.79$), is 1.54 (in the sense $M_G/M_A$), 
the A component 
seems surprisingly large and luminous compared with what it should be on 
the ZAMS; and it ought to be very close to the ZAMS if it is coeval with 
the G component. A possible explanation is that the giant is the merged 
product of a former sub-binary with a mass-ratio of $\ts 0.5$, since this 
could allow the more massive two of the original three components to evolve
at roughly the same rate (Eggleton \& Kiseleva 1996, Eggleton 2006). We 
hope to test shortly the possibility that the appropriate primordial triple 
parameters, from our Monte Carlo model, will give an acceptable number of 
potential  progenitors. Alcock et al. (1999) and Evans et al. (2005) have 
noted that a similar process
might lead to Cepheid binaries of an anomalous character, such as may
be required to reconcile observed Cepheids with the theoretical models
of the Cepheid pulsation phenomenon.

\section*{Acknowledgments}
This work was performed partly under the auspices of the U.S. Department of 
Energy by Lawrence Livermore National Laboratory under Contract 
DE-AC52-07NA27344. We gratefully acknowledge the help
of the Centre des Donn{\'e}es Stellaires  (Strasbourg), and of the
Astronomical Data System.

\bye